\documentclass[a4paper,pdftex,superscriptaddress,preprintnumbers,floatfix,letter,preprint]{elsarticle}
\usepackage{graphicx}

\usepackage{amscd}
\usepackage{amsmath}
\usepackage{amssymb}
\usepackage{bm}
\usepackage{amsthm}
\usepackage{comment}
\usepackage{natbib}

\usepackage{color}

\begin{document}
\bibliographystyle{apsrev}

\title{Minimal Model for Reservoir Computing}

\author{Yuzuru Sato}  
\address{RIES / Department of Mathematics, Hokkaido University,  Kita 12 Nishi 7, Kita-ku, Sapporo, Hokkaido 060-0812, Japan} 
\address{London Mathematical Laboratory, 8 Margravine Gardens, Hammersmith, London W6 8RH, UK}
\author{Miki U. Kobayashi} 
\address{Faculty of Economics, Rissho University, 4-2-16 Osaki, Shinagawa, Tokyo 141-8602, Japan}

\begin{abstract}
 A minimal model for reservoir computing is studied. We demonstrate that a reservoir computer exists that emulates given coupled maps by constructing a modularized network. We describe a possible mechanism for collapses of the emulation in the reservoir computing by introducing a measure of finite scale deviation. Such transitory behaviour is caused by either (i) an escape from a finite-time stagnation near an unstable chaotic set, or (ii) a critical transition driven by the effective parameter drift. Our approach reveals the essential mechanism for reservoir computing and provides insights into the design of reservoir computer for practical applications.
\end{abstract}

\maketitle

\section{Introduction}
Machine learning, including the reservoir computing, has proven useful for model-free and data-driven predictions in a variety of scientific fields in recent years (see \cite{lukovsevivcius2009reservoir,tanaka2019recent} and references therein). In studies on nonlinear phenomena, the reservoir computing is typically applied to large-scale phenomena such as chemical reaction \cite{pathak2018model}, fluid turbulence \cite{nakai_2018, kobayashi_2021}, and climate and environmental dynamics \cite{Coulidaly_2010, mitsui_2021}. 
The purpose of the reservoir computing is to emulate the dynamics $u(t+1)=g(u(t))$ behind time-series data $\{u(t)\}$ using the dynamics $\{r(t)\}$ that is generated by a large recurrent neural network $r(t+1)=f_A(r(t))$ with a network matrix $A$ \cite{Jaeger_2001, Jaeger_2004, Maass_2002}. Assuming that $f_A$ approximates any continuous functions by tuning $A$ when a sufficiently large network is used \cite{cybenko1989approximation}, the trivial solution of this problem is $r(0)=u(0), ~ f_A=g$. 

As opposed to solving the non-convex optimisation problem to approximates $g$ with $f_A$ using a statistical optimisation method such as gradient descent, we may solve a simple convex optimisation problem using linear regression to obtain a projection of the network dynamics $P\{r(t)\}$, which is close to $\{u(t)\}$. The computation cost of the projection $P\{r(t)\}$ is clearly substantially smaller than that of the function approximation of $g$; however, we need to search for an appropriate initial network matrix $A$ for the successful emulation.  
The matrix $A$ is typically provided as a random matrix with a spectral radius $\rho(A)< 1$
\cite{Jaeger_2001} (See Appendix A).
One can search a number of the quenched random dynamics $\{r(t)\}$ by changing the random matrix $A$ to minimise $\lVert P\{r(t)\}-\{u(t)\}\rVert$. 
However, the design principle of such random matrices that satisfy appropriate quality, size, topology, and the order of the search cost is largely unknown. 

As a statistical interpretation, it is pointed out that the reservoir computing is some kind of nonlinear regression \cite{gauthier2021next}. Indeed, a reservoir computing performed by a neural network without hidden layers $f_{O}$ is a simple perceptron, that is equivalent to logistic regression. It is also known that random neural networks also approximate $C^1$ functions arbitrary well \cite{hart2020embedding}, which corresponds to the classical universal approximation theorem \cite{cybenko1989approximation}, and that the non-autonomous framework of reservoir computing is analogous to Takens's embedding theorem, that works even under presence of noise \cite{takens2006detecting,sauer1991embedology,hart2020embedding}. See also Appendix B for the simplest example of a classical delay coordinate embedding performed by a non-autonomous reservoir computer. 

In this study, instead of a Monte Carlo search for the network matrix $A$ and a validation of the quality of the reservoir computing with randomly selected $A$, we construct the smallest possible network that can be included in $A$ to emulate the given dynamical systems. We analyse the behaviour of the constructed dynamical systems, including transitory behaviour that induces the collapse of the emulation \cite{pathak2017using,kong2021machine}. Our approach reveals the essential mechanism for reservoir computing and provides insights into the design of reservoir computer for practical applications. 

\section{Reservoir Computing}
We consider a reservoir computing at time $t$ with $M$-dimensional input neurons $\bm{x}(t)=(x^{(1)}(t),\ldots,x^{(M)}(t))^T$, output neurons $\bm{y}(t)=(y^{(1)}(t),\ldots,y^{(M)}(t))^T$, and a reservoir $\cal{A}$. The reservoir $\cal{A}$ includes $K$ hidden layer neurons $\bm{r}(t)=(r^{(1)}(t),\ldots,r^{(K)}(t))^T$, a $K\times K$ random matrix $A=(a_{ij})$ with a spectral radius $\rho(A)< 1$, and an activation function  $\sigma(\cdot)=\tanh(\cdot)$. The weights for the input and  output are expressed as  $W_{\rm{in}}=(w_{km})$ with $|w_{km}|<1$ and $W_{\rm{out}}=(v_{mk})$, where $m=1,\ldots, M, ~k=1,\ldots, K$, respectively. The bias is given as $\bm{b}=(b_k)$ with $(k=1,\ldots, K)$. We assume that the training data are generated using an $M$-dimensional continuous map $\bm{u}(t)=\bm{g}(\bm{u}(t))$.

In the non-autonomous training phase $t=-N,\ldots, 0$, the entire system dynamics is determinied as follows: 
\begin{equation}
\left\{
\begin{array}{l}
\bm{x}(t)=\bm{u}(t) \\
\bm{y}(t)=W_{\rm{out}}\bm{r}(t)~~~~~~~~~~~~~~~~~~~~~~~~~~(t=-N,\ldots, 0)\\
r^{(k)}(t+1)=(1-\gamma)r^{(k)}(t)+\gamma\tanh\left[(A\bm{r}(t))_k+(W_{\rm{in}}\bm{x}(t))_k+b_k\right]\\
\end{array}
\right.,
\label{eq:training}
\end{equation}
where $\gamma$ denotes the decay rate. 
At t=0, we minimise 
$\lVert \{\bm{y}(t)\}-\{\bm{u}(t)\}\rVert$ based on the generated $\{\bm{r}(t)\}$ to determine the optimal $W_{\rm{out}}=W_{\rm{out}}^*$. With a sufficiently large training data,  linear regression is used to determine the optimal $W_{\rm{out}}^*$. At $t=0$, we replace $\bm{u}(t)$ with $\bm{y}(t)=W_{\rm{out}}^*\bm{r}(t)$, and express 
the system state as  $\bm{x}(0)=\bm{u}(0)=\bm{y}(0)$. 

In the autonomous prediction phase $t=1,2,\ldots$, we obtain
\begin{equation}
\left\{
\begin{array}{l}
\bm{x}(t)=\bm{y}(t)=
W_{\rm{out}}^*\bm{r}(t) ~~~~~~~~~~~~~~~~~~(t=1,2,\ldots)\\
r^{(k)}(t+1)=(1-\gamma)r^{(k)}(t)+\gamma\tanh\left[(A\bm{r}(t))_k+(W_{\rm{in}}\bm{y}(t))_k+b_k\right]
\end{array}
\right., 
\label{eq:prediction}
\end{equation}
and expect the emulation $\{\bm{u}(t)\}\approx\{\bm{y}(t)\}$.

\section{Emulating coupled maps}
\begin{figure}[htbp]
  \begin{center}
    \includegraphics[scale=0.2]{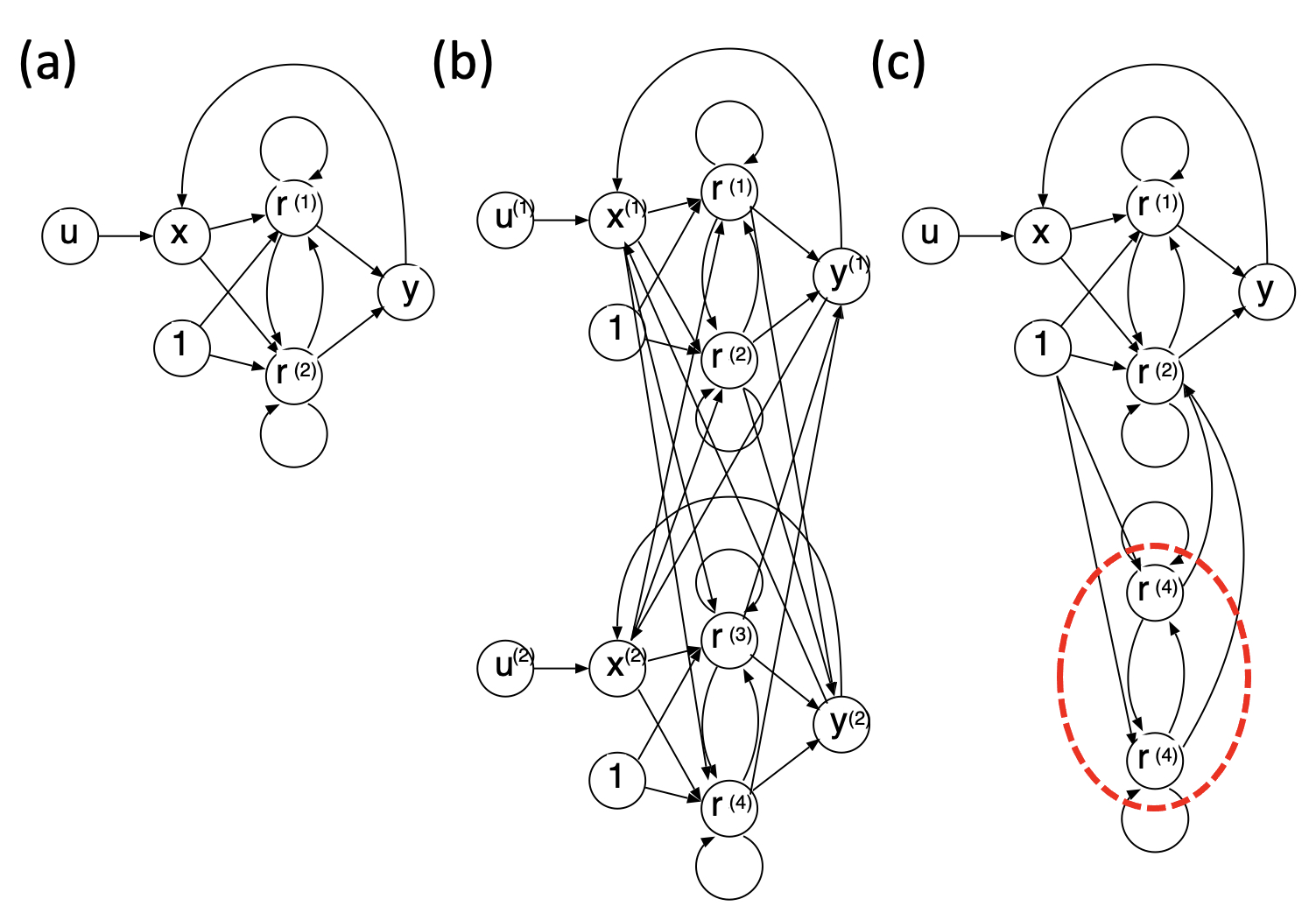}
  \end{center}
  \caption{(a) Model of $K=2$ reservoir  for  one-dimensional map,  (b) model of $K=4$ reservoir for $2$-coupled maps, and (c) model of emulation collapse induced by excess neurons. The dotted circle indicates the collective excess neurons.}
  \label{model}
\end{figure}

\begin{figure}[htbp]
  \begin{center}    \includegraphics[scale=0.13]{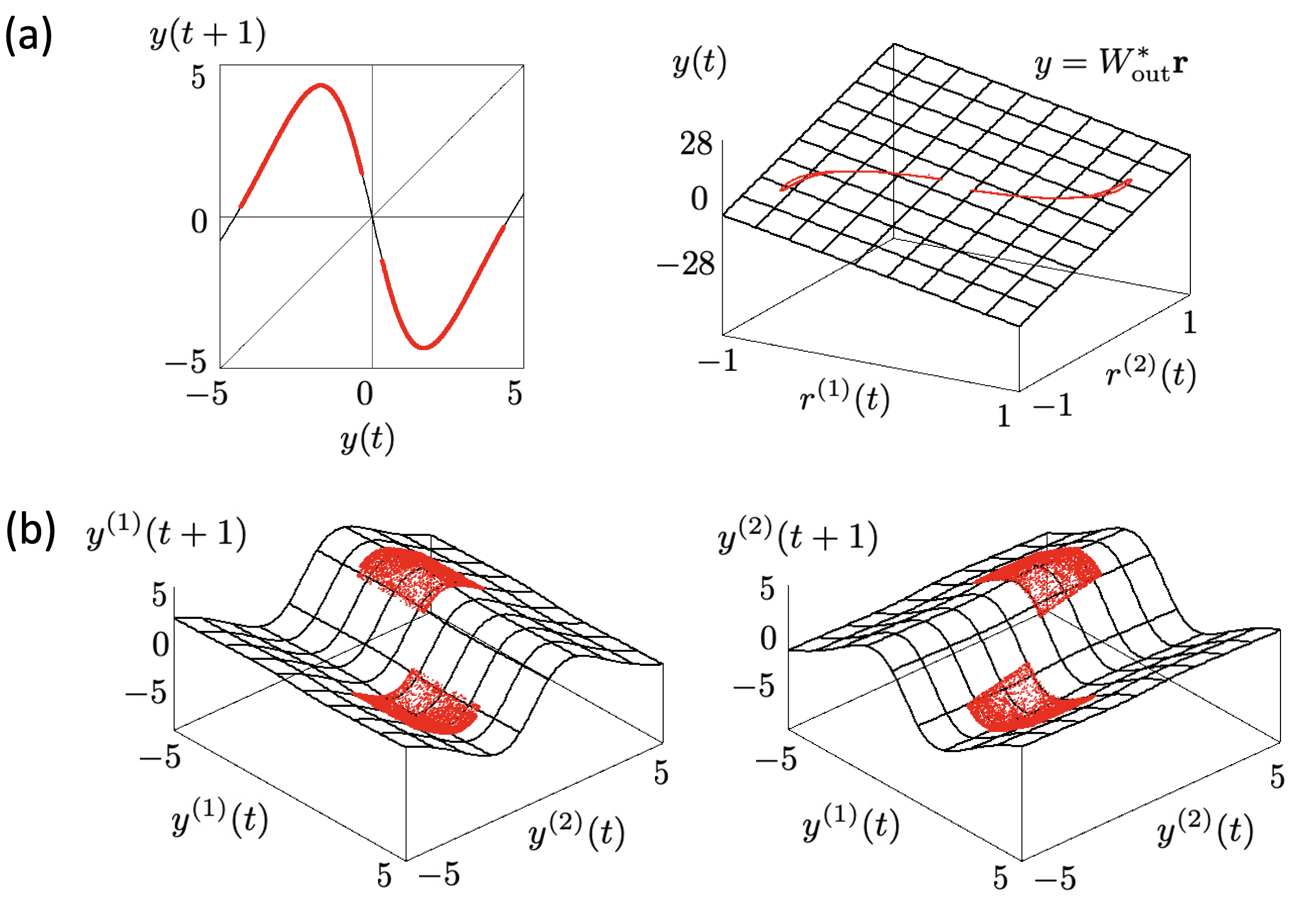}
  \end{center}
  \caption{(a) Left:  ~ the map provided in Eq. (\ref{eq:map})  and  return plots of $\{y(t)\}$ generated by constructed reservoir with $K=2$, and right: ~ the attractor in $(r^{1)},r^{(2)},y)$, and the plane  $y=W_{\rm{out}^*}\bm{r}$. 
  (b) The map provided in Eq. (\ref{eq:map2}) and  return plots of $\{\bm{y}(t)\}$ generated by constructed reservoir with $K=4$. The parameters are $\gamma=0.9, ~~\beta_1=0.75, ~~\beta_2=0.25, ~~\alpha_1=\alpha_2=0, ~~v^*_1=-12, ~~v_2^*=16, \epsilon=0.05$, and $\mu_1=\mu_2=\mu_3=\mu_4=0.01$. }
  \label{fig:1d2dmap}
\end{figure}

To construct a minimal example, we assume that the training data are  generated by a class of one-dimensional maps given as 
\begin{equation}
u(t+1)=(1-\gamma)u(t)+\gamma \sum_{k=1,2} v_k^*\tanh\left[\beta_k u(t)+\alpha_k\right].
\label{eq:map}
\end{equation} 
Eq. (\ref{eq:map}) shows  chaotic behaviour
in a broad range of  parameters. We adopt $\gamma=0.9, ~~\beta_1=0.75, ~~\beta_2=0.25, ~~\alpha_1=\alpha_2=0, ~~v^*_1=-12$, and $v_2^*=16$ as an example of a generator of chaotic training data. We emulate the dynamics of Eq. (\ref{eq:map}) using a reservoir $\cal{B}$ with $K=2$ (Fig. \ref{model} (a)) to set 
\begin{equation}
B=\begin{pmatrix}
\mu_1 v_1^* & \mu_1 v_2^*\\
\mu_2 v_1^* & \mu_2 v_2^*\\
    \end{pmatrix}, ~~0<\mu_1,\mu_2\ll 1, ~\rho(B)< 1, 
    \label{2dres}
\end{equation}
\begin{equation}
W_{\rm{in}}=\begin{bmatrix}
\beta_1-\mu_1\\
\beta_2-\mu_2\\
\end{bmatrix}, 
~~W_{\rm{out}}=\begin{bmatrix}
v_1 & v_2\\
\end{bmatrix}, 
~~\bm{b}=\begin{bmatrix}
\alpha_1\\
\alpha_2\\
\end{bmatrix}. 
\label{2dweight}
\end{equation}
As $\mu_1, \mu_2$ are free parameters, we may set $B$ satisfying the condition $\rho(B)<1$. 
Following linear regression with a sufficiently large training data $\{u(t)\}_{t=-N,\ldots,0}$, 
$W_{\rm{out}}$ is optimised to \begin{eqnarray}
    W_{\rm{out}}^*=[v_1^* ~ v_2^*] 
\end{eqnarray} 
and we obtain  
\begin{equation}
\left\{
\begin{array}{l}
x(t)=y(t)=v_1^*r^{(1)}(t)+v_2^*r^{(2)}(t)\\
r^{(1)}(t+1)
=(1-\gamma)r^{(1)}(t)+ \gamma\tanh\left[\beta_1 y(t)+\alpha_1\right]\\
r^{(2)}(t+1)
=(1-\gamma)r^{(2)}(t)+\gamma\tanh\left[\beta_2 y(t)+\alpha_2\right] \\
\end{array}
\right.,
\label{eq:k2res}
\end{equation}
resulting in
\begin{equation}
y(t+1)=(1-\gamma)y(t)+\gamma \sum_{k=1,2} v_k^*\tanh\left[\beta_k y(t)+\alpha_k\right],
\label{eq:resmap}
\end{equation}
which is equivalent to Eq. (\ref{eq:map}). Recall that at $t=0$ we obtain  $x(0)=u(0)=y(0)$, and the dynamics is on $x=y=W_{\rm{out}}^*\bm{r}=v_1^*r^{(1)}+v_2^*r^{(2)}$, which may include the generalised synchronisation manifold in the reservoir $\cal{B}$ for the input $\{u(t)\}$ \cite{lu2018attractor,lymburn2019reservoir,kocarev1996generalized}. 
Thus, we achieve a perfect emulation $\{y(t)\}=\{u(t)\}$ for $t=0,1,2,\ldots$ by using the $K=2$ reservoir $\cal{B}$. We confirm that the above emulation is achieved in numerical experiments with linear regression (Fig. \ref{fig:1d2dmap} (a)). Many other possible constructions are available for given one-dimensional maps.

Next, we emulate a class of coupled maps, as follows:   
\begin{equation}
\left\{
\begin{array}{l}
u^{(1)}(t+1)=(1-\epsilon)f(u^{(1)}(t))+\epsilon f(u^{(2)}(t))\\
u^{(2)}(t+1)=(1-\epsilon)f(u^{(2)}(t))+\epsilon f(u^{(1)}(t))\\
f(u)=(1-\gamma)u+\gamma \sum_{k=1,2} v_k^*\tanh\left[\beta_k u+\alpha_k\right]\\
\end{array}
\right..
\label{eq:map2}
\end{equation}
To emulate the dynamics of Eq. (\ref{eq:map2}) using a reservoir $\cal{C}$ with $K=4$ (Fig. \ref{model} (b)), we set 
\begin{equation}
\begin{array}{l}
C=\begin{pmatrix}
\mu_1 (1-\epsilon)v_1^* & \mu_1 (1-\epsilon)v_2^* & \mu_1\epsilon v_1^* & \mu_1\epsilon v_2^*\\
\mu_2 (1-\epsilon)v_1^* & \mu_2 (1-\epsilon)v_2^*&\mu_2\epsilon v_1^*  & \mu_2\epsilon v_2^*\\
\mu_3\epsilon v_1^* & \mu_3\epsilon v_2^* &\mu_3 (1-\epsilon)v_1^* & \mu_3 (1-\epsilon)v_2^*\\
\mu_4\epsilon v_1^* & \mu_4\epsilon v_2^*& \mu_4 (1-\epsilon)v_1^* & \mu_4 (1-\epsilon)v_2^*\\
\end{pmatrix}, \\
\\
~~~~~~~~~~0<\mu_1,\mu_2,\mu_3,\mu_4\ll 1, ~~\rho(C)< 1 \\
    \end{array},
\end{equation}
\begin{equation}
W_{\rm{in}}=\begin{bmatrix}
\beta_1-\mu_1&0\\
\beta_2-\mu_2&0\\
0&\beta_1-\mu_3\\
0&\beta_2-\mu_4\\
\end{bmatrix},
~~W_{\rm{out}}=\begin{bmatrix}
v_{11}&v_{12}&v_{13}&v_{14}\\
v_{21}&v_{22}&v_{23}&v_{24}\\
\end{bmatrix}, 
~~\bm{b}=\begin{bmatrix}
\alpha_1\\
\alpha_2\\
\alpha_1\\
\alpha_2\\
\end{bmatrix}. 
\end{equation}
Following linear regression, 
$W_{\rm{our}}$ is optimised to 
\begin{equation}
W_{\rm{out}}^*=\begin{bmatrix}
(1-\epsilon) v_1^*&(1-\epsilon) v_2^*&\epsilon  v_1^*&\epsilon  v_2^*\\
\epsilon  v_1^*&\epsilon  v_2^*&(1-\epsilon) v_1^*&(1-\epsilon)v_2^*\\
\end{bmatrix}, 
\end{equation}
and we obtain  
\begin{equation}
\left\{
\begin{array}{l}
x^{(1)}(t)=y^{(1)}(t)=(W_{\rm{out}}^*\bm{r}(t))_1\\
x^{(2)}(t)=y^{(2)}(t)=(W_{\rm{out}}^*\bm{r}(t))_2\\
r^{(1)}(t+1)=(1-\gamma)r^{(1)}(t)+\gamma\tanh\left[\beta_1 y^{(1)}(t)+\alpha_1\right]\\
r^{(2)}(t+1)=(1-\gamma)r^{(2)}(t)+\gamma\tanh\left[\beta_2y^{(1)}(t)+\alpha_2\right] \\
r^{(3)}(t+1)=(1-\gamma)r^{(3)}(t)+\gamma\tanh\left[\beta_1 y^{(2)}(t)+\alpha_1\right]\\
r^{(4)}(t+1)=(1-\gamma)r^{(4)}(t)+\gamma\tanh\left[\beta_2 y^{(2)}(t)+\alpha_2\right] \\
\end{array}
\right.,
\end{equation}
resulting in 
\begin{equation}
\left\{
\begin{array}{l}
y^{(1)}(t+1)=(1-\epsilon)f(y^{(1)}(t))+\epsilon f(y^{(2)}(t))\\
y^{(2)}(t+1)=(1-\epsilon)f(y^{(2)}(t))+\epsilon f(y^{(1)}(t))\\
f(y)=(1-\gamma)y+\gamma \sum_{k=1,2}v_k^*\tanh\left[\beta_ky+\alpha_k\right]\\
\end{array}
\right., 
\end{equation}
which is equivalent to Eq. (\ref{eq:map2}). 
Noting that $\bm{x}(0)=\bm{u}(0)=\bm{y}(0)$, and the dynamics is on $\bm{x}=\bm{y}=W_{\rm{out}}^*\bm{r}$, we achieve a perfect emulation $\{\bm{y}(t)\}=\{\bm{u}(t)\}$. We confirm that the above emulation is achieved in numerical experiments with linear regression (Fig. \ref{fig:1d2dmap} (b)). 

In general, according to the property of neural networks as universal function approximators, with sufficiently many neurons, we may emulate any one-dimensional maps 
\begin{equation}
    u(t+1)=(1-\gamma)u(t)+\gamma h(u(t)),
\end{equation}
where $h$ denotes an  arbitrary continuous functions. In practice, it is impossible to learn structurally unstable dynamical systems with finite data, however, in machine learning, input data always includes observational noise and/or finite size fluctuation, and thus function approximabilities is typically discussed in theory.
For a large $K$, the projected dynamics $\{y(t)\}$ of a reservoir with $K$ neurons can be expressed as follows: 
\begin{equation}
\begin{array}{l}
y(t+1)=(1-\gamma)y(t)+\gamma \sum_{k=1}^Kv_k^*\tanh\left[\beta_k y(t)+\alpha_k\right]
\\
~~\stackrel{K\to\infty}{\longrightarrow} ~~y(t+1)=(1-\gamma)y(t)+\gamma h(y(t)).\\
\end{array}
\end{equation}
For example, the projection $\{y(t)\}$ may emulate the logistic map $y(t+1)=a-y(t)^2$ when $h(y)= -(1-\gamma)y+a-y^2$. Similarly, we can emulate arbitrary $M$-coupled maps with a linear coupling. To do that we use an $M$-dimensional projected dynamics of a reservoir with $KM$ neurons and a network matrix that consists of $M$ blocks of the size $K$. For example, with the construction in Eq. (\ref{eq:map2}) where $f(y) = a-y^2$, and the adequately optimised $W_{\rm{out}}^*$, the $M$-dimensional projected dynamics $\{\bm{y}(t)\}$ can emulate the dynamics of $M$-globally coupled logistic map $y^{(m)}(t+1)=(1-\epsilon)f(y^{(m)}(t))+\frac{\epsilon}{M}\sum_{i=1}^M f(y^{(i)}(t))$. This construction for coupled maps partially explains the successful emulation of PDEs by reservoir computers \cite{pathak2017using, pathak2018model}. 
Using a reservoir with a sufficiently large network, it is certainly possible to emulate the dynamics of a wide class of nonlinear PDEs, which are implemented on computers in the form of coupled maps. However, the essential problem is rather why a finite-size reservoir computer with a randomly selected network matrix efficiently emulates such dynamical systems with large degrees of freedom.

\section{Emulation in large networks}
Although the matrix $B$ or $C$ may be a part of a larger network matrix $A$,  in certain cases in large networks, the system may degenerate and the effective dynamics may depend only on a smaller part of the network represented by $B$ or $C$. We consider such degeneration to the reservoir $\cal{B}$. 

In general, some of neurons may not contribute to the emulation. We refer to a neuron with index $k$, where $w_{km}=v_{mk}=0$ holds for all $m$,  
as an "excess neuron."  Assuming that $r^{(k)} ~(k=3,\ldots,N)$ are excess neurons,  
and that their dynamics vanishes as $r^{(k)}\to 0$ when without the bias, the degeneration to the perfect emulation is achieved. We set a reservoir $\cal{A}$ with $K$ neurons, as follows: 
\begin{equation}
A=\begin{pmatrix}
B&R_1\\
O&R_2\\    
\end{pmatrix}, ~~\rho(A)< 1, 
\end{equation}
\begin{equation}
W_{\rm{in}}=\begin{bmatrix}
\beta_1-\mu_1\\
\beta_2-\mu_2\\
0\\
\vdots\\
0\\
\end{bmatrix},
~~W_{\rm{out}}=\begin{bmatrix}
v_{1}\\
v_{2}\\
v_3\\
\vdots \\
v_K\\
\end{bmatrix}^T, 
~~\bm{b}=\begin{bmatrix}
\alpha_1\\
\alpha_2\\
0\\
\vdots\\
0\\
\end{bmatrix}. 
\end{equation}
When $R_1=O$ and $R_2$ is a random matrix, $W_{\rm{out}}$ is optimised to $W_{\rm{out}}^*=\begin{bmatrix}
v_{1}^* & v_{2}^* & 0 &\cdots & 0\\
\end{bmatrix}$, and Eq. (\ref{eq:map}) is emulated. In other cases, 
the projected dynamics $\{\bm{y}(t)\}$ may deviate from the perfect emulation  
by the perturbations from the collective excess neurons, which we discuss later in the paper.

The system may also degenerate by synchronisation of neurons. 
For example, we set a reservoir $\cal{D}$ as follows; 
\begin{equation}
\begin{array}{l}
D=\begin{pmatrix}
\mu_1 \frac{v_1^*}{2} & \mu_1 \frac{v_2^*}{2}&\mu_1 \frac{v_1^*}{2} & \mu_1 \frac{v_2^*}{2}\\
\mu_2 \frac{v_1^*}{2} & \mu_2 \frac{v_2^*}{2}&\mu_2 \frac{v_1^*}{2} & \mu_2 \frac{v_2^*}{2}\\
\mu_3 \frac{v_1^*}{2} & \mu_3 \frac{v_2^*}{2}&\mu_3 \frac{v_1^*}{2} & \mu_3 \frac{v_2^*}{2}\\
\mu_4 \frac{v_1^*}{2} & \mu_4 \frac{v_2^*}{2}&\mu_4 \frac{v_1^*}{2} & \mu_4 \frac{v_2^*}{2}\\
    \end{pmatrix}, 
~~0<\mu_1,\mu_2,\mu_3,\mu_4\ll 1, ~\rho(D)< 1 \\
\end{array},
    \label{2dres2}
\end{equation}
\begin{equation}
W_{\rm{in}}=\begin{bmatrix}
\beta_1-\mu_1\\
\beta_2-\mu_2\\
\beta_1-\mu_3\\
\beta_2-\mu_4\\
\end{bmatrix}, 
~~W_{\rm{out}}^*=\begin{bmatrix}
v_1^* & v_2^* &v_3^* & v_4^*\\
\end{bmatrix}, 
~~\bm{b}=\begin{bmatrix}
\alpha_1\\
\alpha_2\\
\alpha_1\\
\alpha_2\\
\end{bmatrix}. 
\label{4dweight}
\end{equation}
and obtain 
\begin{equation}
\left\{
\begin{array}{l}
\bm{x}(t)=\bm{y}(t)=W_{\rm{out}}^*\bm{r}(t)\\
r^{(1)}(t+1)
=(1-\gamma)r^{(1)}(t)+ \gamma\tanh\left[\beta_1 y(t)+\alpha_1\right]\\
r^{(2)}(t+1)
=(1-\gamma)r^{(2)}(t)+\gamma\tanh\left[\beta_2 y(t)+\alpha_2\right] \\
r^{(3)}(t+1)
=(1-\gamma)r^{(3)}(t)+ \gamma\tanh\left[\beta_1 y(t)+\alpha_1\right]\\
r^{(4)}(t+1)
=(1-\gamma)r^{(4)}(t)+\gamma\tanh\left[\beta_2y(t)+\alpha_2\right] \\
\end{array}
\right..
\label{eq:sync}
\end{equation}
The dynamics of $(r^{(3)}, r^{(4)})$ in Eq. (\ref{eq:sync}) 
is eventually synchronised to those of $(r^{(1)},r^{(2)})$ because 
\begin{equation}
\left\{
\begin{array}{l}
r^{(1)}(t)-r^{(3)}(t)=(1-\gamma)(r^{(1)}(t)-r^{(3)}(t))\\
r^{(2)}(t)-r^{(4)}(t)=(1-\gamma)(r^{(2)}(t)-r^{(4)}(t))
\end{array}
\right..
\label{eq:sync2}
\end{equation}
As a result, the dynamics of $y(t)$ prefectly emulates Eq. (\ref{eq:map}). In large matrices, when the $K$-dimensional vector $\bm{r}$ is degenerated to a $2$-dimensional vector $(r^{(1)},r^{(2)})$ based on the synchronization to $2$ clusters $r^{(1)}=r^{(i)}, r^{(2)}=r^{(j)}, ~(i\neq j, ~3\le i,j\le K)$, again, the perfect emulation of Eq. (\ref{eq:map}) is achieved. Many other constructions are available for the degeneration by synchronisation. 

While in gradient descent learning in multi-layer perceptrons, the degeneration of the network weights may cause vanishing gradient and decelerate learning process  \cite{fukumizu2000local,sato2022noise}, the degeneration of the hidden layer dynamics in the reservoir computing may enhance precise emulations. 

\section{Effectively successful emulation}
To evaluate emulation of one-dimensional maps 
  $u(t+1)=g(u(t))\in[s_1,s_2]$, 
  we introduce a finite scale deviation $D_{t_0}(T,\delta)$ in a time window $t_0-T\le t\le t_0+T$ 
  between the input data $\{u(t)\}$ generated by the dynamical system $g$, and the emulation $\{y(t)\}$ by a reservoir computer, based on an observation within a finite time $2T$ and a finite spatial resolution $\delta$. 
First, we set a partition $\{\Delta_i\}$ 
with $m$ bins of the size $\delta$ for the interval $[s_1,s_2]$ as $\Delta_i=[s_1+(i-1)\delta,s_1+i\delta], ~~ (i=1,\ldots,m, ~m\delta=|s_2-s_1|)$,
and observe the dynamics $\{y(t)\}$ visiting the $i$th small interval $\Delta_i$. 
When we observe the dynamics in $\Delta_i$ for $k^{(i)}>0$ times at $t=\tau_1^{(i)},\ldots,\tau_k^{(i)}$, we take the center of mass of the tuple  $(y(\tau^{(i)}),y(\tau^{(i)}+1))$ over the generated data, 
given as
\begin{equation}
(\tilde{y}^{(i)}, \tilde{y}^{(i)}{}')
= \left(
\frac{y(\tau_1^{(i)})+\cdots y(\tau_{k^{(i)}}^{(i)})}{k^{(i)}}, ~
\frac{y(\tau_1^{(i)}+1)+\cdots y(\tau_{k^{(i)}}^{(i)}+1)}{k^{(i)}}\right), 
\end{equation}
where  $y(\tau_1^{(i)}),\ldots,y(\tau_{k^{(i)}}^{(i)})\in \Delta_i$. The tuple $(\tilde{y}^{(i)}, \tilde{y}^{(i)}{}')$ 
is the representative of successive observations of the dynamics in $\Delta_i$. When the dynamics does not visit $\Delta_i$, that is $k^{(i)}=0$, we define $(\tilde{y}^{(i)},\tilde{y}^{(i)}{}')=(s_1+(i-\frac{1}{2})\delta, 0)$. 
Then the finite scale deviation measured in a time window $t_0-T\le t \le t_0+T$ is defined as 
  \begin{equation}
D_{t_0}(T,\delta)=
\frac1m\sum_{i=1}^{m}(g(\tilde{y}^{(i)})-\tilde{y}^{(i)}{}')^2,
\end{equation}
where the limit $
  D_{t_0}^{\infty}=\lim_{T\to \infty}\lim_{\delta\to 0}D_{t_0}(T,\delta)$,
approximating the deviation between $\{(\tilde{y}^{(i)}, g(\tilde{y}^{(i)}))\}$ and  $\{(\tilde{y}^{(i)},\tilde{y}^{(i)}{}') ~~(i=1,\ldots,m)\}$ in a given finite scale. For the perfect emulation, we have $D_t^{\infty}=0$ for all $t$. We may estimate $g(\tilde{y}^{(i)})$ with the input data by a linear interpolation in case that we do not know $g$.  
If a criterion $D_{t_0}(T,\delta)<d$ is satisfied, we say that the simulation is {\it effectively successful}, in a time window $t_0-T\le t \le t_0+T$,  with a deviation $d$, in a finite time $2T$, and in a  finite spatial resolution $\delta$. 

\begin{figure}[htbp]
  \begin{center}
    \includegraphics[scale=0.18]{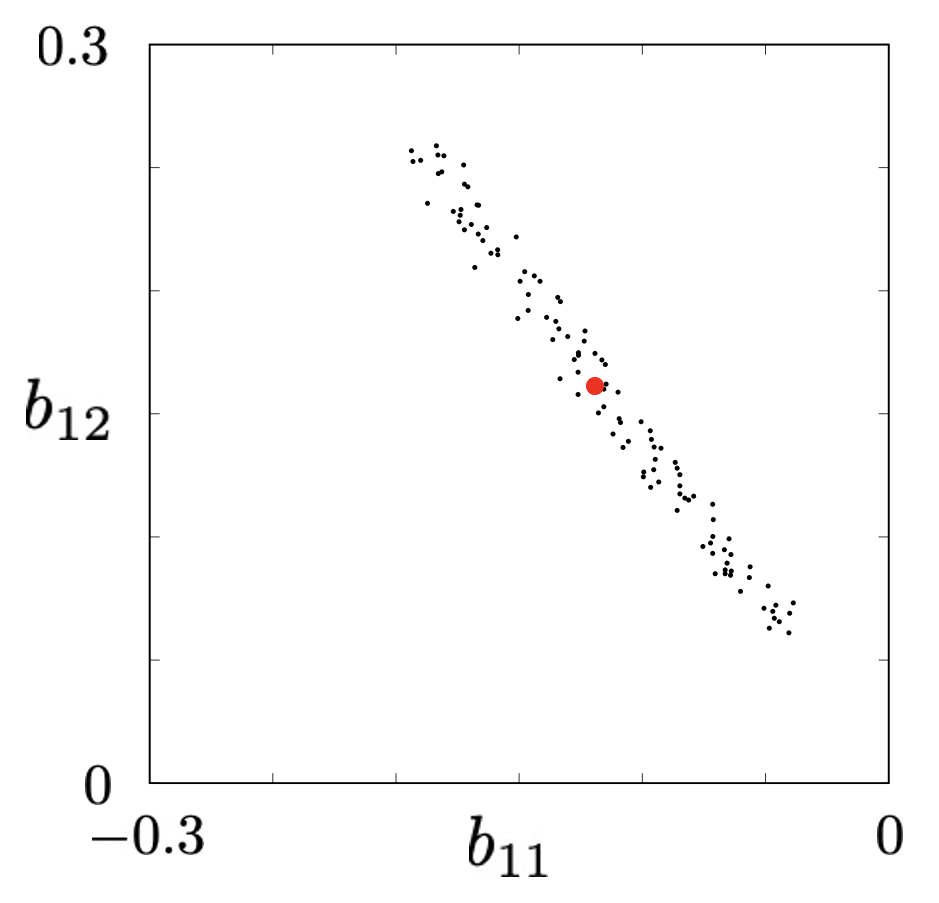}
  \end{center}
  \caption{The perfect emulation surrounded by effectively successful  emulation: The black points plotted near a section $(b_{21},b_{22})=(-0.12,0.16)$,  indicate the effectively successful emulation after $N=10^4$ training phase with a criterion $D_{t_0}(T,\delta)<d$. The red point indicates the perfect emulation given by $(b_{11},b_{12})=(-1.2,1.6)$. Around the perfect emulation point, there is a region consists of the black points corresponding to effectively successful emulation. The parameters are given as $t_0=1500$, $T=1500$, $\delta=10^{-3}$, and $d=5 \times 10^{-4}$. }
  \label{fig:effemu}
\end{figure}

As an example, we compute the deviation $D_{t_0}(T,\delta)$ with $t_0=1500$, $T=1500$, $\delta=10^{-3}$, and $d=5 \times 10^{-4}$  
for the reservoir computing given in Eqs. (3)-(6). The parameters are given as $\mu_1=\mu_2=0.01$, $\beta_1=0.75, \beta_2=0.25$, $\alpha_1=\alpha_2=0$, and the matrix $B$ is given as the following random matrix;
\begin{equation}
B=
\begin{pmatrix}
b_{11} & b_{12}\\
b_{21} & b_{22}\\ 
\end{pmatrix}
=
\begin{pmatrix}
-0.12 & 0.16\\
-0.12 & 0.16\\ 
\end{pmatrix}
+
\begin{pmatrix}
\xi_{11} & \xi_{12}\\
\xi_{21} & \xi_{22}\\  \end{pmatrix},
\end{equation}
where $\xi_{ij}\in [-0.8,0.8]$ randomly chosen from uniform distribution. In FIG \ref{fig:effemu}, the black points plotted near a section $(b_{21},b_{22})=(-0.12,0.16)$, indicate the effectively successful emulation after $N=10^4$ training phase with a criterion $D_{t_0}(T,\delta)<d$. The red point indicates the perfect emulation,  with $D_t^{\infty}=0$ for all $t$, given by   $(b_{11},b_{12})=(-1.2,1.6)$, which induces the correct estimation of the original model with  $(v_1^*,v_2^*)=(-12,16)$. Around the perfect emulation point, there is a region consists of the black points corresponding to effectively successful emulation. 

Note that the region consists of the black dots in FIG \ref{fig:effemu}  includes "nearly"  successful emulation achieved by transient chaos, due to our criterion based on the finite space-time deviation. In the region far from the perfect emulation point, the estimated dynamical system is possibly after a bifurcation to a periodic window. In such cases, although the dynamics in the autonomous prediction phase starts with $y(0)=u(0)$, where $u(0)$ is on the attractor of the original dynamics $\{u(t)\}$, the initial point $y(0)$ is not on the original attractor, but is very close to the chaotic saddle of the projected dynamics $\{y(t)\}$ and may show a very long transient chaos similar to the original dynamics. 

\section{Collapse of emulation}

When excess neurons exist, the dynamics $\{\bm{y}(t)\}$ in the subspace $\bm{x}=\bm{y}=W_{\rm{out}}^*\bm{r}$ can be perturbed, and the emulation may collapse in the long run. We create a model for the collapse of the emulation in  Eq. (\ref{eq:map}) using a reservoir with $K=4$ and   
\begin{equation}
\begin{array}{l}
E=\begin{pmatrix}
\mu_1 v_1^* & \mu_1 v_2^* & 0 & 0\\
\mu_2 v_1^* & \mu_2 v_2^* & q_1 & q_2\\
0 & 0 & \nu_1 q_1 & \nu_1 q_2\\
0 & 0 & \nu_2 q_1 & \nu_2 q_2\\
    \end{pmatrix}, 
    ~~0<\mu_1,\mu_2\ll 1, ~\rho(E)< 1 \\
    \end{array},
\end{equation}
\begin{equation}
W_{\rm{in}}=\begin{bmatrix}
\beta_1-\mu_1\\
\beta_2-\mu_2\\
0\\
0\\
\end{bmatrix},  
 ~~W_{\rm{out}}^*=\begin{bmatrix}
v_{1}^*&v_{2}^*&0&0\\
\end{bmatrix}, 
 ~~\bm{b}=\begin{bmatrix}
\alpha_1\\
\alpha_2\\
\alpha_3\\
\alpha_4\\
\end{bmatrix}.
\end{equation}
For the entire system dynamics, we obtain  
\begin{equation}
\left\{
\begin{array}{l}
x(t)=y(t)= v_1^*r^{(1)}(t)+ v_2^*r^{(2)}(t)\\
\eta(t+1)=q_1r^{(3)}(t)+q_2r^{(4)}(t)\\
r^{(1)}(t+1)=(1-\gamma)r^{(1)}(t)+\gamma\tanh\left[
\beta_1 y(t) +\alpha_1\right] \\
r^{(2)}(t+1)=(1-\gamma)r^{(2)}(t)+\gamma\tanh\left[
\beta_2 y(t)+\alpha_2 +\eta(t)\right] \\
r^{(3)}(t+1)=(1-\gamma)r^{(3)}(t)+\gamma\tanh\left[
\nu_1\eta(t)+\alpha_3\right] \\
r^{(4)}(t+1)=(1-\gamma)r^{(4)}(t)+\gamma\tanh\left[
\nu_2\eta(t)+\alpha_4\right] \\
\end{array}
\right.,
\end{equation}
\begin{equation}
\begin{array}{l}
y(t+1)=(1-\gamma)y(t)+\gamma (v_1^*\tanh\left[\beta_1 y(t)+\alpha_1\right]+v_2^*\tanh\left[\beta_2 y(t)+\alpha_2+\eta(t))\right]\\
\end{array},
\label{eq:perturbedmap}
\end{equation}
\begin{equation}
\eta(t+1)=(1-\gamma)\eta(t)+\gamma \left( q_1\tanh\left[\nu_1 \eta(t)+\alpha_{3}\right]+q_2\tanh\left[\nu_2 \eta(t)+\alpha_{4}\right]\right). 
\label{eq:tangency}
\end{equation}

When $\eta(t)=0$, 
linear regression yields $W=W_{\rm{out}}^*$ and 
 Eq. (\ref{eq:perturbedmap}) is equivalent to Eq. (\ref{eq:map}).  Thus, the collectives of excess neurons provides  external perturbations to the projected dynamics $\{y(t)\}$. We present two examples of  the collapse of the emulation in these types. 
 
\begin{figure}[htbp]
  \begin{center}
\includegraphics[scale=0.25]
{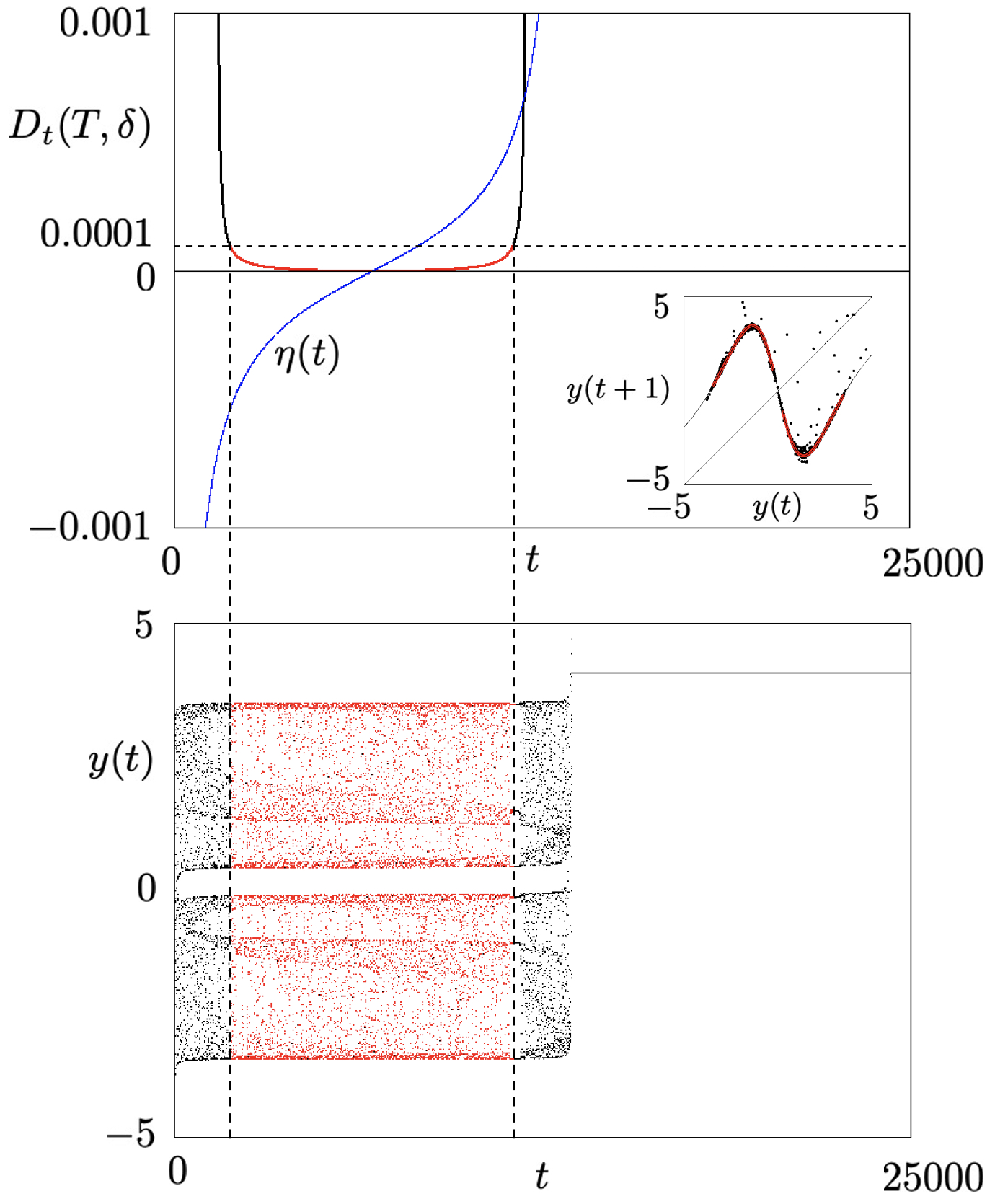}
\end{center}
  \caption{ The emulation collapse caused by a tangency:  
  The origin is almost neutrally stable, and the fixed point $\eta(t)=\eta^*\simeq 3.773542$ is stable in Eq. (\ref{eq:tangency}). The dynamics of $\eta(t)$ and the deviation  $D_t(T,\delta)$, where $T=1500, ~\delta=10^{-3}$, is depicted in the blue and the black line, respectively, in the upper panel. The red part of $D_t$ is below the threshold $D_t<d=0.0001$.  
  In the lower panel, the time series of $y(t)$ is depicted in black dots, and, in the red part, the emulation is effectively successful. The inset of the upper panel is the return plot of $\{y(t)\}$, and the red part is the finite time return plot of $\{y(t)\}$ during effectively successful emulation. The parameters are $\gamma=0.9$, $q_1=-12, q_2=16, \beta_1=\nu_1=0.75, \beta_2=0.25, \nu_2=0.5, \alpha_1=\alpha_2=0, \alpha_3=\tanh^{-1}(2/3)-10^{-8}$, and $\alpha_4=\tanh^{-1}(1/2)$. 
  }
  \label{fig:tangency}
\end{figure}

 In the first example, we set $\gamma=0.9, q_1=-12, q_2=16, \beta_1=\nu_1=0.75$, $\beta_2=0.25$, $\nu_2=0.5$, and $\alpha_1=\alpha_2=0$. When $\alpha_3=\tanh^{-1}(2/3)$ and $\alpha_4=\tanh^{-1}(1/2)$, Eq. (\ref{eq:tangency}) has a tangency at the origin and a stable fixed point $\eta=\eta^*\simeq 3.773542$. 
In this case, the perfect emulation can be performed as a neutral solution on the centre manifold $\eta(t)=0$. The chaotic saddle is approached when we start at an initial point in the effective basin defined by $\eta(0)<0$ and $\eta(0)\notin(-2.197527,-0.985279)$; otherwise, the dynamics converges to $\eta(t)=\eta^*$ and the emulation fails. In general, if the system is at the onset of a bifurcation and have a narrow channel at the stagnation point, the dynamics stays near the stagnation point for a long time. The duration time near the stagnation point typically follows a power law, and can be arbitrarily longer. If the dynamics of $\eta(t)$ is near the stagnation point, 
and the resulting waiting time is longer than the training time $N$, the projected dynamics $\{y(t)\}$ can emulate $\{u(t)\}$ for a finite time. In particular, when $\eta(t)=0$, the perfect emulation is achieved.  However, in the long run, the dynamics of $\eta(t)$ escapes from the neighbourhood of the stagnation point and converges to the stable fixed point $\eta(t)=\eta^*\simeq 3.773542$. Thus, the emulation is effectively successful for a finite time but eventually collapses (FIG \ref{fig:tangency}).

\begin{figure}[htbp]
  \begin{center}
\includegraphics[scale=0.25]{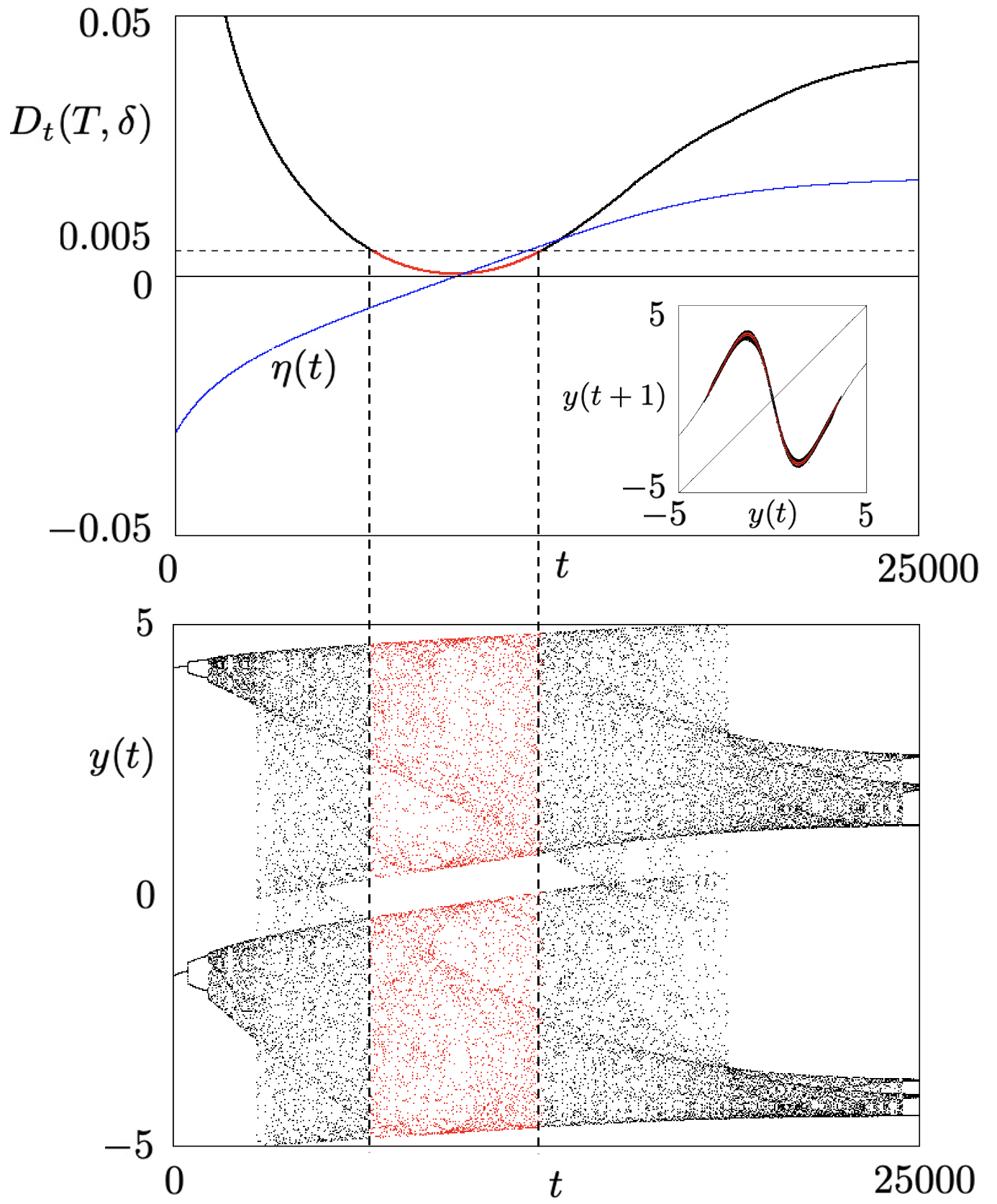}
  \end{center}
  \caption{
  The  emulation collapse caused by a parameter drift:  
  There is an almost neutrally stable narrow channel near the origin, and the fixed point $\eta(t)=\eta^*\simeq 0.019036$ is stable in Eq. (\ref{eq:ct}). The dynamics of $\eta(t)$ and the deviation $D_t(T,\delta)$, where  $T=1500, ~\delta=10^{-3}$ is depicted in the blue and the black line, respectively, in the upper panel. The red part of $D_t$ is below the threshold $D_t<d=0.005$. In the lower panel,  
  the time series of $y(t)$ is depicted in black dots, and in the red part, the emulation is effectively successful. The inset of the upper panel is the return plot of $\{y(t)\}$, and the red part is the finite time return plot of $\{y(t)\}$ during effectively successful emulation. 
  The parameters are  $\gamma=0.9$, $q_1=1, \nu_1=1$, and $\alpha_3=2.3\times 10^{-6}$.
  }
\label{fig:ct}
\end{figure}

The second example is based on only one excess neuron with $\eta(t+1)=q_1r^{(3)}(t)$. We set $\gamma=0.9, q_1=1, \nu_1=1$, and $\alpha_3=2.3\times 10^{-6}$, 
 and obtain a hyperbolic tangent map
\begin{equation}
\eta(t+1)=(1-\gamma)\eta(t)+\gamma q_1\tanh\left[\nu_1\eta(t)+\alpha_3\right],
\label{eq:ct}
\end{equation}
where $\eta(0)=-0.03$.  In this case, the perfect emulation can be performed on an unstable chaotic set. The dynamics of Eq. (\ref{eq:ct}) does not have a tangency but has an almost neutrally stable narrow channel near the origin, and has a stable fixed point $\eta(t)=\eta^*\simeq 0.019036$.  The origin can be approached when we start at $\eta(0)<0$; otherwise, the dynamics converges to the stable fixed point $\eta(t)=\eta^*$ and the emulation fails.  A very slow uniform motion $\eta(t)\simeq \gamma\tanh(\alpha_3) t\simeq 10^{-6}t$ emerges near the origin, which works as a slow parameter drift in Eq. (\ref{eq:perturbedmap}). Thus, considering a situation that other neurons in the large network work as external noise, the critical transition may occur with the parameter drift and the external noise \cite{scheffer2009early, kong2021machine}. We observe orbits that are close to the original dynamics for a finite time, however, before $\eta$ reaches to the stable fixed point $\eta(t)=\eta^*\simeq 0.019036$, it may arrive at the tipping point of Eq. (\ref{eq:perturbedmap})  and the emulation collapses by the critical transition (FIG \ref{fig:ct}).

In these two scenarios, the initial matrices and the initial conditions of $\bm{r}(0)$ have a positive measures in the parameter and state space. Thus, the above phenomena can be observed in a quenched random dynamical systems with a random matrix $A$ and with an initial condition $\bm{r}(0)$.

\section{Conclusion}

We have demonstrated that a reservoir computer exists that emulates given coupled maps by constructing a modularised network. We have proposed  a possible mechanism for the collapses of the emulation in reservoir computing  by introducing a measure of finite scale deviation. Such transitory behaviour is caused by either (i) a finite-time stagnation near an unstable chaotic set or (ii) a critical transition by the effective parameter drift. The essential problem in reservoir computing is determining why a finite-size reservoir computer with a randomly selected  network can efficiently emulate dynamical systems with large degrees of freedom. 
Our approaches provides a minimal model for understanding reservoir computing, thereby providing better insights into the design of reservoir computer for practical applications. Problems in the quenched random dynamical systems, such as bifurcations, generalised synchronisation, and various types of the emulation collapse, will be studied elsewhere. 





\section*{Acknowledgements}

We acknowledge H. Suetani (Oita University) for useful discussion and comments.
The research leading to these results has been partially funded by the Grant in 
Aid for Scientific Research (B) No. 21H01002, JSPS, Japan, and the London Mathematical Laboratory External Fellowship, United Kingdom. 

\begin{appendix}
\section{Echo state property}

The echo state property is a characteristic of a given reservoir dynamics with a network matrix $A$, where the reservoir dynamics converge to the same reservoir behavior as time tends to infinity, regardless of the choice of initial reservoir states \cite{yildiz2012re}. As an empirical measure of the echo state property, the condition of the spectral radius of the network matrix $\rho(A)< 1$ has been discussed. However, in our construction, we can set arbitrary spectral radii for perfect emulations of any coupled maps. In general, if multiple attractors exist, the echo state property does not hold \cite{ceni2020echo}. 

\section{Delay coordinate embedding by reservoir computing}

Assuming that the training data is provided as one-dimensional time series $\{u(t)\}$ and the attractor of the dynamics generating the training data $\bm{y}(t+1)=\bm{g}(\bm{y}(t))$ is a compact manifold of dimension $d$, the non-autonomous reservoir computer given by Eq. (\ref{eq:training}) with  $2d+1$ output neurons, and with a smooth activation function is embedding by the embedding theorem \cite{takens2006detecting, sauer1991embedology}.  
Both embedding and the reservoir computing unfold the training data into a high-dimensional space (or a reservoir), and take a projection of the high-dimensional structure. To see that with a simple example, we construct a $K=3$ reservoir computer $\cal{Q}$ with $1$-dimensional input neurons, $3$-dimensional output neurons, that embeds $1$-dimensional time series to $3$-dimensional state space. We assume that the training data are provided as $\{u(t)\}$ that is generated by a dynamical system whose effective dimension of stationary attractor is less than $3$. When the training data $\{u(t)\}$ is scaled to the dynamics very close to the origin, the activation function $\sigma(x)=\tanh(x)$ is an almost identical function as $\tanh(x)\approx x ~(x\approx 0)$. We set $\gamma=1$, and 

 \begin{equation}
Q=\begin{pmatrix}
0&0&0\\
1&0&0\\
0&1&0\\
\end{pmatrix}, 
~~W_{\rm{in}}=\begin{bmatrix}
1\\
0\\
0\\
\end{bmatrix}, 
~~W_{\rm{out}}^*=\begin{bmatrix}
1&0&0\\
0&1&0\\
0&0&1\\
\end{bmatrix}, 
~~\bm{b}=\begin{bmatrix}
0\\
0\\
0\\
\end{bmatrix}, 
\label{3demb}
\end{equation}
and obtain
\begin{equation}
\left\{
\begin{array}{l}
  x(t)=u(t)\\
  \bm{y}(t)=W_{\rm{out}}^*\bm{r}(t)=\bm{r}(t)\\
r^{(1)}(t+1)=\tanh(x(t))\approx x(t)\\
r^{(2)}(t+1)=\tanh(r^{(1)}(t))\approx r^{(1)}(t)\\
r^{(3)}(t+1)=\tanh(r^{(2)}(t))\approx r^{(2)}(t)\\
\end{array}
\right..
\end{equation}
Thus, $\bm{y}(t)$ is the classical delay coordinate embedding of the training data $\{u(t)\}$ 
\begin{equation}
\left\{
    \begin{array}{l}
    y^{(1)}(t+3)=r^{(1)}(t+3)\approx  x(t+2)=u(t+2)\\
    y^{(2)}(t+3)=r^{(2)}(t+3)\approx  x(t+1)=u(t+1)\\
    y^{(3)}(t+3)=r^{(3)}(t+3)\approx  x(t)=u(t)\\
    \end{array}
    \right.
\end{equation}
via a reservoir $\cal{Q}$. See \cite{hart2020embedding} for the general embedding theorem for non-autonomous echo state networks.

\end{appendix}

\bibliography{rc}

\end{document}